\documentclass[preprint2]{aastex} 

\shorttitle{\textit{r}-Process in Anisotropic Neutrino Winds}

\shortauthors{Wanajo}

\begin{document}

\title{The {\boldmath \textit{r}}-Process in the Proto-Neutron-Star
Winds with Anisotropic Neutrino Emission}

\author{Shinya Wanajo}

\affil{Department of Astronomy, School of Science,
   University of Tokyo, Bunkyo-ku, Tokyo, 113-8654, Japan;
   wanajo@astron.s.u-tokyo.ac.jp}

\begin{abstract}
The astrophysical origin of the \textit{r}-process nuclei is still
unknown. Even the most promising scenario, the neutrino-driven winds
from a nascent neutron star, encounters severe difficulties in
obtaining requisite entropy and short dynamic timescale for the
\textit{r}-process. In this study, the effect of anisotropy in
neutrino emission from a proto-neutron star surface is examined with
semi-analytic neutrino-driven wind models. The increase of neutrino
number density in the wind owing to the anisotropy is modeled
schematically by enhancing the \textit{effective} neutrino
luminosity. It is shown that the neutrino heating rate from
neutrino-antineutrino pair annihilation into electron-positron pairs
can significantly increase owing to the anisotropy and play a dominant
role for the heating of wind material. A factor of five increase in
the effective neutrino luminosity results in $50 \%$ higher entropy
and a factor of ten shorter dynamic timescale owing to this enhanced
neutrino heating. The nucleosynthesis calculations show that this
change is enough for the robust \textit{r}-process, producing the
third abundance peak ($A = 195$) and beyond. Future multi-dimensional
studies with accurate neutrino transport will be needed if such
anisotropy relevant for the current scenario (more than a factor of a
few) is realized during the wind phase ($\sim 1-10\, \textrm{s}$).
\end{abstract}


\section{Introduction}
The astrophysical site of the rapid-neutron-capture nucleosynthesis
(\textit{r}-process), which accounts for about half of nuclei heavier
than iron, has been a long-standing mystery. During the last decade,
the neutrino-heated ejecta from a nascent neutron star
\citep[neutrino-driven winds,][]{Woos94} has been considered to be the
most promising astrophysical site for the \textit{r}-process. Previous
studies show, however, severe problems in obtaining requisite high
entropy and short dynamic timescale for the production of heavy
\textit{r}-process nuclei \citep{Qian96, Otsu00, Sumi00, Wana01,
Thom01}. The general relativistic effect for a very compact
proto-neutron star \citep[e.g., the mass of $2.0\, M_\odot$ with the
radius of 10~km,][]{Otsu00, Wana01} or a magnetar-like magnetic field
strength \citep{Thom03, Suzu05} have been invoked to increase entropy
and reduce the dynamic timescale of the winds. It is questionable,
however, if such physical conditions can be the general requirements
for the \textit{r}-process nucleosynthesis.

\citet{Qian96} have suggested that an additional energy input to the
neutrino-driven wind at between 1.5 and 3 times the neutron star
radius is efficient to increase entropy and reduce dynamic timescale
of the wind material. In this \textit{Letter}, it is shown that strong
anisotropy in neutrino emission from the proto-neutron star,
\textit{if it exists}, acts as this extra energy source and helps the
\textit{r}-process. The neutrino-driven wind model with spherically
symmetric, steady outflow approximation is used to obtain the wind
trajectories (\S~2). A sudden increase of neutrino number density in
winds owing to anisotropic neutrino emission is modeled by enhancing
the neutrino luminosity. Nucleosynthesis calculations with the
obtained thermodynamic trajectories are performed to demonstrate this
effect (\S~3). Finally, a possible origin of this anisotropy in
neutrino emission and some implications of this study are discussed
(\S~4).

\section{Wind Models with Anisotropic Neutrino Emission}

The wind trajectories in this study are obtained using the
semi-analytic, general relativistic model of neutrino-driven winds
explored in \citet{Otsu00} and \citet{Wana01, Wana02}. In this model,
the system is treated as time stationary and spherically symmetric,
and the radius of the neutron star is assumed to be the same as that
of the neutrino sphere. The heating source that drives matter from the
neutrino sphere is due to neutrino interactions. Heating is due to
$\nu_e$ and $\bar{\nu}_e$ capture on free nucleons ($\dot q_{\nu N}$),
neutrino scattering by electrons and positrons ($\dot q_{\nu e}$), and
neutrino-antineutrino pair annihilation into electron-positron pairs
($\dot q_{\nu \nu}$). Cooling is due to electron and positron capture
on free nucleons ($\dot q_{e N}$) and electron-positron pair
annihilation into neutrino-antineutrino pairs ($\dot q_{e e}$). The
rms average neutrino energies are taken to be 10, 20, and 30~MeV, for
electron, anti-electron, and the other flavors of neutrinos,
respectively. The mass ejection rate at the neutrino sphere $\dot M$
is determined so that the wind becomes supersonic through the sonic
point.

The mass and radius of the neutron star are taken to be $M = 1.4\,
M_\odot$ and $R = 10\, \mathrm{km}$, respectively. The neutrino
luminosity of one specific flavor is assumed to be the same for all
other flavors, which is taken to be a constant value $L_\nu = 1 \times
10^{51}\, \mathrm{ergs}\, \mathrm{s}^{-1}$. Note that the assumption
of a constant $L_\nu$ is reasonable, since the crossing time of a wind
over the heating region ($< 30\, \mathrm{km}$, see Fig.~2) is short
enough, $\sim 0.1\, \mathrm{s}$, compared to the decay timescale of
$L_\nu$ \citep[a few seconds, e.g.,][]{Woos94}. As explored in
previous studies, this \textit{typical} choice of parameter set (with
isotropic neutrino emission) results in insufficient physical
conditions (i.e., low entropy and long dynamic timescale) for the
production of heavy \textit{r}-process nuclei \citep[e.g.,][]{Wana01}.

In this study, anisotropy in neutrino emission is modeled
schematically as follows. Given there is substantially higher neutrino
emission from the ``hot spot'', which is marked by the point P$_1$
($\mathrm{OP}_1 = R$) in Figure~1. At the point P$_0$ ($\mathrm{OP}_0
= R$) nearby P$_1$, the ejection of matter is due to the local (lower)
isotropic neutrino emission around P$_0$. The matter suddenly sees a
substantially larger number of neutrinos when passing through the
point P$_2$. Note that neutrino emission at an arbitrary point on the
neutrino sphere (e.g., P$_0$ or P$_1$) is assumed to be isotropic in
all directions (i.e., the local neutrino flux is \textit{not} radial)
as in \citet{Otsu00}. This sudden increase of the neutrino number
density at P$_2$ is approximated by a jump of the neutrino luminosity
from the original value $L_\nu = 1 \times 10^{51}\, \mathrm{ergs}\,
\mathrm{s}^{-1}$ for $R < r < R_2$ to the \textit{effective}
luminosity $L_{\nu 2}$ for $r \ge R_2$, where $r$ is the distance from
the center O and $R_2 = \mathrm{OP}_2$.

The wind models considered in this study are listed in the first
column of Table~1, where $R_2$ (second column) and $L_{\nu 2}$ (third
column) are taken to be 10, 12, 15, and 20~km, and 1.0, 2.0, 3.0, 4.0,
and 5.0 in units of $10^{51}\, \mathrm{ergs}\, \mathrm{s}^{-1}$. The
resulting net heating rate ($\dot q$; \textit{top panel}) for models
A1, A5, B5, C5, and D5 and each heating/cooling rate (\textit{bottom
panel}) for models A5 and B5 are shown in Figure~2, as functions of
$r$. Note that A1-A5 are \textit{isotropic} wind models (i.e., $R_2 =
R$).

For isotropic winds (A1-A5 in Table~1), the higher $L_{\nu 2}$ ($=
L_\nu$ in these cases) results in shorter dynamic timescale
$\tau_\mathrm{dyn}$ ($\equiv |\rho/(d\rho/dt)|_{T = 0.5\,
\mathrm{MeV}}$) and \textit{lower} asymptotic (i.e., maximum) entropy
$s$. This shows that the increased $\dot q$ (see A1 and A5 in Fig.~2,
\textit{top panel}) is consumed to drive more matter (i.e., higher
$\dot M$ as can be seen in Table~1) from the neutron star surface with
faster velocity, rather than to increase entropy. In contrast, for
anisotropic models, an increase of $L_{\nu 2}$ (for $r \ge R_2$) is
quite efficient \textit{both} to increase entropy and to reduce
dynamic timescale (Table~1). The reason is that the matter has been
already lifted with low $L_\nu\, (= 1 \times 10^{51}\, \mathrm{ergs}\,
\mathrm{s}^{-1} < L_{\nu 2})$ and thus with small $\dot M$. Therefore,
the density (and temperature) at arbitrary $r$ is significantly small
compared to the corresponding isotropic wind. This can be seen in the
5th (and 6th) column in Table~1, which lists the density $\rho_{13}$
(and temperature $T_{13}$) at $r = 13\, \mathrm{km}$ (see about one
order difference in $\rho_{13}$ for A5 and B5).

For isotropic wind models, the five times greater neutrino luminosity
simply results in the increase of $\dot q$ with the same factor (A1
and A5 in Fig.~2, \textit{top panel}). This does not hold, however,
for anisotropic wind models. For model B5, the maximum $\dot q$ is as
twice large as that for model A5 (with the same $L_{\nu 2}$), and more
than 10 times larger than that for model A1 (with the same
$L_\nu$). This can be explained as follows. As shown in Figure~2
(\textit{bottom panel}), for isotropic winds (\textit{dashed lines};
A5), the heating is mainly due to $\dot q_{\nu N}$ and $\dot q_{\nu
e}$, while $\dot q_{\nu \nu}$ plays only a minor role. In contrast,
for anisotropic winds (\textit{solid lines}; B5), the neutrino pair
annihilation $\dot q_{\nu \nu}$ plays a crucial role, whose peak (at
$r \approx 13\, \mathrm{km}$) is a factor of seven higher than that in
A5. This effect can be clearly seen in Figure~2 (\textit{top panel}),
in which the case without an increase of $\dot q_{\nu \nu}$
(model~B5a) and with an increase of $\dot q_{\nu \nu}$ only
(model~B5b) are compared (see also Table~1).

This is due to the difference of $\rho$ and $T$ dependences in these
heating terms. For a fixed set of $r$, $Y_e$, $L_\nu$ and neutrino
mean energies, these heating rates are related to $\rho$ and $T$ such
as $\dot q_{\nu N} = \textrm{constant}$, $\dot q_{\nu e} \propto T^4
\rho^{-1}$, and $\dot q_{\nu \nu} \propto \rho^{-1}$ \citep{Qian96,
Otsu00}. As a result, $\dot q_{\nu N}$ in B5 (Fig.~2, \textit{bottom
panel}) closely follows that in A5 for $r > R_2$, which is independent
of $\rho$ and $T$. As can be seen in Table~1, a reduction in $\rho$
owing to low $L_\nu$ in B5 (compared to that in A5) is accompanied
with a reduction in $T$. As a consequence, $\dot q_{\nu e}$ in B5 is
lower than that in A5 even for $r > R_2$. However, $\dot q_{\nu \nu}$
is not dependent on $T$ but is inversely proportional to $\rho$ (i.e.,
proportional to the number of neutrinos per volume), which becomes
significantly high owing to the decreasing $\rho$ for $r > R_2$
(Table~1). Note that the cooling terms ($\dot q_{e N} \propto T^6$ and
$\dot q_{e e} \propto T^9 \rho^{-1}$) quickly decay with increasing
$r$ and have negligible effects by the anisotropy (Fig.~2).

As can be seen in the above numerical experiments, the strong
anisotropy in neutrino emission can be an additional energy source
pointed out by \citet{Qian96}. However, this mechanism may work only
for $r < 1.5\, R$, which is rather closer to the neutrino sphere than
the suggested range ($1.5 < r/R < 3$) by \citet{Qian96}. In the
current study, the effects of increasing entropy and accelerating wind
are prominent for the wind closer to the hot spot, in particular for
$R \approx 12\, \mathrm{km}$ (Table~1), at which $\dot q$ maximizes
(Fig.~2). The effect of anisotropic neutrino emission becomes less
important for a more distant wind (e.g., $R_2 = 20\, \mathrm{km}$),
where the neutrino heating has mostly ceased (Fig.~2). For a fixed
$R_2$, the effect is more significant for higher $L_{\nu 2}$ as can be
seen in Table~1. For model B5, the entropy is about $50 \%$ higher
($180\, N_A\, k$) and the dynamic timescale is about a factor of ten
shorter ($1.65\, \mathrm{ms}$) than those in the isotropic model A1
with the same $L_\nu$ ($s = 117\, N_A\, k$ and $\tau_\mathrm{dyn} =
14.1\, \mathrm{ms}$).

Note that the purely parametric examinations explored in this section
should be regarded as only qualitative ones. For instance, the ``hot
spot'' is not necessary a point as illustrated in Figure~1. It is
conceivable that the area with strong neutrino emission has some
distribution on the neutrino sphere. Moreover, the configuration
should become multi-dimensional soon after the wind material passes
the point P$_2$ in Figure~1, which is treated within the framework of
a spherical wind model in the current study. More realistically, the
wind matter deviates from the radial to the direction of OP$_2$ in
Figure~1. This may moderate the acceleration of wind and increase the
heating duration. Hence, the current results may overestimate the
reduction of dynamic timescale and underestimate the increase of
entropy. It is difficult to estimate the net effect to the
nucleosynthesis from these modifications. Obviously, a
multi-dimensional approach will be needed to quantitatively estimate
the effects of the anisotropy.

\section{Nucleosynthesis in Winds}

Adopting the wind trajectories discussed in \S~2 for the physical
conditions, the nucleosynthetic yields are obtained by solving an
extensive nuclear reaction network. The network consists of 6300
species between the proton and neutron drip lines \citep[for more
detail, see][]{Wana06}. Neutrino-induced reactions and nuclear fission
are not considered in the current study. Each calculation is initiated
when the temperature decreases to $T_9 = 9$ (where $T_9 \equiv
T/10^9\, \mathrm{K}$). The initial compositions are given by $X_n = 1
- Y_{e}$ and $X_p = Y_{e}$, respectively, where $X_n$ and $X_p$ are
the mass fractions of neutrons and protons, and $Y_{e}$ is the initial
electron fraction (number of proton per nucleon) at $T_9 = 9$. In this
study, $Y_{e}$ is taken to be 0.4, according to the core-collapse
simulation in \citet[][at $L_\nu \approx 1 \times 10^{51}\,
\mathrm{ergs}\, \mathrm{s}^{-1}$]{Woos94}. As in \citet{Wana02}, the
temperature and density are set to be constant when $T_9$ decreases to
$1.0$, in order to mimic the effect of the slower outgoing ejecta
behind the shock.

The nucleosynthesis results for models B2-B5, C2-C5, and D2-D5
(Table~1) are shown in Figure~3, as a function of atomic mass number.
For anisotropic wind models with $R_2 = 12\, \mathrm{km}$ (B2-B5), the
effect of anisotropic neutrino emission is evident. A factor of three
or four increase in $L_{\nu 2}$ (B3 and B4 in Table~1) leads to $s
\approx 150-160\, N_A\, k$ and $\tau_\mathrm{dyn} \approx 3-4\,
\mathrm{ms}$, resulting in the \textit{r}-process nucleosynthesis
(Fig.~3). For model B5, the high entropy ($= 180\, N_A\, k$) and short
dynamic timescale ($= 1.65\, \mathrm{ms}$) of the wind drive the
nuclear matter to the actinide region. The neutron-to-seed abundance
ratio at the beginning of the \textit{r}-process, defined as $T_9 =
2.5$, is $Y_n/Y_h = 176$ and the final averaged mass number of heavy
nuclei with $Z > 2$ is $\langle A_h \rangle = 230$ (Table~1). For the
models with $R_2 = 15\, \mathrm{km}$ (C2-C5), the \textit{r}-process
still takes place when $L_{\nu 2}$ is four or five times higher than
$L_\nu$ (models C4 and C5). For $R_2 = 20\, \mathrm{km}$ (D2-D5), the
effect of anisotropic neutrino emission is not important and the
nucleosynthesis results are not significantly different from the
isotropic cases (A1-A5).

\section{Implications}

In this \textit{Letter}, the effects of anisotropy in neutrino
emission for the \textit{r}-process nucleosynthesis in
proto-neutron-star winds were examined, using the spherically
symmetric, steady outflow model of neutrino-driven winds. It was shown
that strong anisotropy, \textit{if it exists}, can be an additional
energy source \citep{Qian96} to heat the wind material. A factor of
four or five enhancement in \textit{effective} neutrino luminosity
results in the significant increase of entropy and shortening of
dynamic timescale of outgoing neutrino-heated ejecta. This is mainly
due to the \textit{boosted} neutrino heating from annihilation of
neutrino-antineutrino pairs into electron-positron pairs as a result
of anisotropic neutrino emission. This provides the physical condition
suitable for the robust \textit{r}-process, producing the third
abundance peak ($A = 195$) and beyond.

It is conceivable that asymmetric neutrino emission can be associated
with the anisotropic matter distribution near the neutrino sphere. As
an example, \citet{Kota03} suggested that the non-spherical neutrino
sphere owing to rapid rotation leads to anisotropic neutrino heating
with the pole-to-equator ratio of a few to more than 10. This may
result in strong contrast in neutrino emission on the neutrino sphere,
which forms an \textit{effective} ``hot spot'' around the rotational
axis. A recent work with more sophisticated neutrino-transport scheme
by \citet{Wald05} showed, however, that the pole-to-equator flux ratio
is at most a factor of two, even for a rather rapidly rotating
core. This is a consequence that the radiation field is smoothened by
the many neutrino sources above the neutrino sphere (e.g., convective
bubbles) at the early phase ($< 1\, \textrm{s}$ after core
bounce). Nevertheless, all the convective bubbles are evacuated during
the late wind phase ($\sim 10\, \textrm{s}$) and a strong contrast of
neutrino flux might form on the neutrino sphere for a rapidly rotating
core.

Another possibility of anisotropic neutrino emission might be due a
global fluid instabilities of neutrino-heated matter as observed in
multi-dimensional hydrodynamic simulations. Recent works have shown
that hydrodynamic instabilities can lead to low-mode ($l = 1$ in terms
of an expansion in spherical harmonics of order $l$) oscillation of
the convective fluid flow in the neutrino-heated layer behind the
shock \citep[e.g.,][]{Sche06, Bura06b, Burr06}. The presence of such a
low convective mode results in the pair of a single outflow and a
narrow accretion flow that creates the ``hot spot'' on the neutron
star surface. It should be noted, however, that the two-dimensional
simulations by \citet{Sche06} showed that the anisotropy of the
accretion luminosity owing to this flow appears to be only a few
percent (at least during the early phase up to $\sim 1\, \textrm{s}$
after core bounce). A future investigation relevant for the wind phase
($\sim 1-10\, \textrm{s}$) will be needed to examine the degree of
anisotropic neutrino emission from such an accretion flow.

Given one of the above (or another unknown) mechanism works, a
constraint for the \textit{r}-process may be obtained from the
condition that creates the ``hot spot'' owing to, e.g., rapid rotation
or long lasting accretion flow. It is conceivable that only a limited
fraction of supernovae create the ``host spot'' relevant for the
current scenario (e.g., rapid rotators or less-energetic supernovae
that form the long lasting accretion flow). This can be a reasonable
explanation for that the spectroscopic analysis of extremely
metal-poor stars and Galactic chemical evolution study imply only a
limited fraction of core-collapse supernovae undergo the
\textit{r}-process nucleosynthesis \citep{Ishi99, Ishi04}.

The implications in this study must be tested by future
multi-dimensional simulations of core-collapse supernovae for long
duration ($\sim 10\, \mathrm{s}$) with accurate neutrino
transport. Systematic calculations of nucleosynthesis with such
hydrodynamic trajectories will be also needed to investigate the
contribution to the Galactic chemical evolution of \textit{r}-process
nuclei.

\acknowledgements

I would like to acknowledge H. -Th. Janka for helpful discussions. I
also acknowledge the contributions of an anonymous referee, which led
to clarification of a number of points in the original
manuscript. This work was supported in part by a Grant-in-Aid for
Scientific Research (17740108) from the Ministry of Education,
Culture, Sports, Science, and Technology of Japan.


\begin{deluxetable}{cccccccccc}
\tablecaption{Model Parameters and Results}
\tablewidth{0pt}
\tablehead{
\colhead{} &
\colhead{$R_2$} &
\colhead{$L_{\nu 2}$ (10$^{51}$} &
\colhead{$\dot M$ (10$^{-6}$} &
\colhead{$\rho_{13}$ (10$^7$} &
\colhead{$T_{13}$ (10$^{10}$} &
\colhead{$s$} &
\colhead{$\tau_\mathrm{dyn}$} &
\colhead{} &
\colhead{}
\\
\colhead{Model} &
\colhead{(km)} &
\colhead{ergs s$^{-1}$)} &
\colhead{$M_\odot\, \mathrm{s}^{-1})$} &
\colhead{g cm$^{-3}$)} &
\colhead{K)} &
\colhead{$(N_A\, k)$} &
\colhead{(ms)} &
\colhead{$Y_n/Y_h$} &
\colhead{$\langle A_h \rangle$}
}
\startdata
A1  & 10 & 1.0 & 3.86 & 3.49 & 2.06 & 117 & 14.1 & 6.01 & 107 \\
A2  & 10 & 2.0 & 13.3 & 5.98 & 2.35 & 103 & 7.20 & 8.16 & 109 \\
A3  & 10 & 3.0 & 27.4 & 8.10 & 2.54 & 95.5 & 4.99 & 6.02 & 107 \\
A4  & 10 & 4.0 & 46.0 & 9.75 & 2.66 & 90.7 & 3.90 & 6.02 & 107 \\
A5  & 10 & 5.0 & 68.7 & 11.9 & 2.80 & 87.0 & 3.26 & 10.3 & 111 \\
B2  & 12 & 2.0 & 4.88 & 2.68 & 1.88 & 131 & 7.45 & 18.0 & 118 \\
B3  & 12 & 3.0 & 5.84 & 2.08 & 1.73 & 145 & 4.44 & 38.9 & 135 \\
B4  & 12 & 4.0 & 6.84 & 1.65 & 1.61 & 161 & 2.72 & 80.7 & 170 \\
B5  & 12 & 5.0 & 7.93 & 1.15 & 1.46 & 180 & 1.65 & 176 & 230 \\
B5a\tablenotemark{a} & 12 & 5.0 & 5.84 & 2.07 & 1.74 & 147 & 4.37 & 39.8 & 136 \\
B5b\tablenotemark{b} & 12 & 5.0 & 6.89 & 1.64 & 1.60 & 162 & 2.61 & 86.9 & 174 \\
C2  & 15 & 2.0 & 4.15 & 3.37 & 1.99 & 127 & 9.76 & 12.1 & 113 \\
C3  & 15 & 3.0 & 4.41 & 3.25 & 1.94 & 136 & 7.08 & 20.9 & 120 \\
C4  & 15 & 4.0 & 4.68 & 3.25 & 1.90 & 147 & 5.15 & 34.7 & 131 \\
C5  & 15 & 5.0 & 4.97 & 3.10 & 1.84 & 159 & 3.69 & 58.3 & 153 \\
D2  & 20 & 2.0 & 3.93 & 3.45 & 2.04 & 122 & 11.9 & 8.43 & 109 \\
D3  & 20 & 3.0 & 4.00 & 3.53 & 2.04 & 127 & 10.2 & 11.3 & 112 \\
D4  & 20 & 4.0 & 4.08 & 3.50 & 2.02 & 132 & 8.69 & 14.9 & 115 \\
D5  & 20 & 5.0 & 4.15 & 3.48 & 2.01 & 137 & 7.43 & 19.3 & 119 \\
\enddata
\tablenotetext{a}{Same as B5, but without enhancement of $\dot q_{\nu \nu}$ for $r > R_2$.}
\tablenotetext{b}{Same as B5, but with enhancement of only $\dot q_{\nu \nu}$ for $r > R_2$.}
\end{deluxetable}


\begin{figure}
\plotone{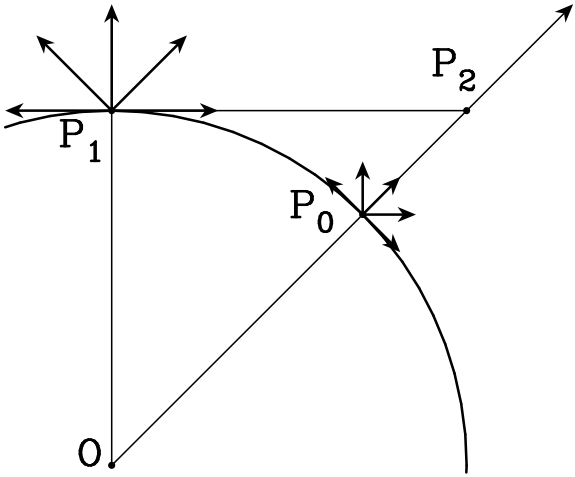}
\caption{Illustration of asymmetric neutrino emission. O is the center
of the neutron star. Strong neutrino emission from the ``hot spot''
near the point P$_1$ on the neutrino sphere is assumed, otherwise
being isotropic. A wind blowing from a nearby point P$_0$ with the
(weaker) isotropic neutrino emission ($L_\nu$) suddenly see a larger
number of neutrinos ($L_{\nu 2}$) when passing P$_2$.}
\end{figure}


\begin{figure}
\plotone{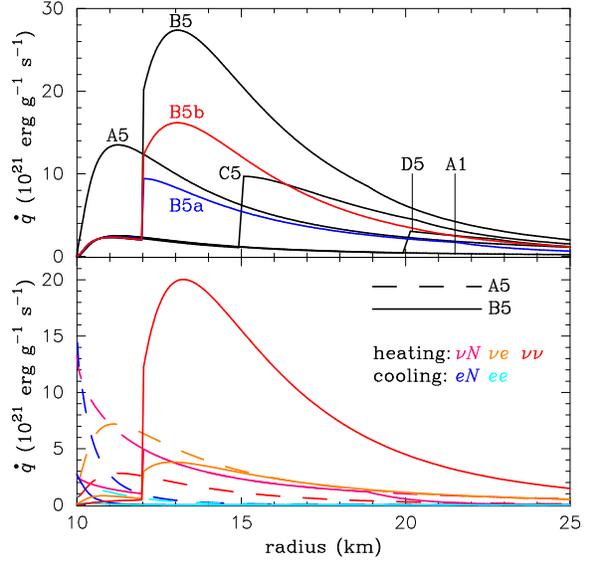}
\caption{\textit{Top}: Net neutrino heating rates for wind models A1,
A5, B5, B5a, B5b, C5, and D5 listed in Table~1, as a function of
$r$. Jump of the heating rate at $r = R_2$ is due to the sudden
increase of effective neutrino luminosity from $L_\nu$ to $L_{\nu 2}$.
\textit{Bottom}: Heating ($\dot q_{\nu N}$, $\dot q_{\nu e}$, and
$\dot q_{\nu \nu}$) and cooling ($\dot q_{e N}$ and $\dot q_{e e}$)
rates as functions of $r$. Dashed and solid lines are for wind models
A5 and B5 (listed in Table~1), respectively.}
\end{figure}

\clearpage

\begin{figure}
\epsscale{2.0} 
\plotone{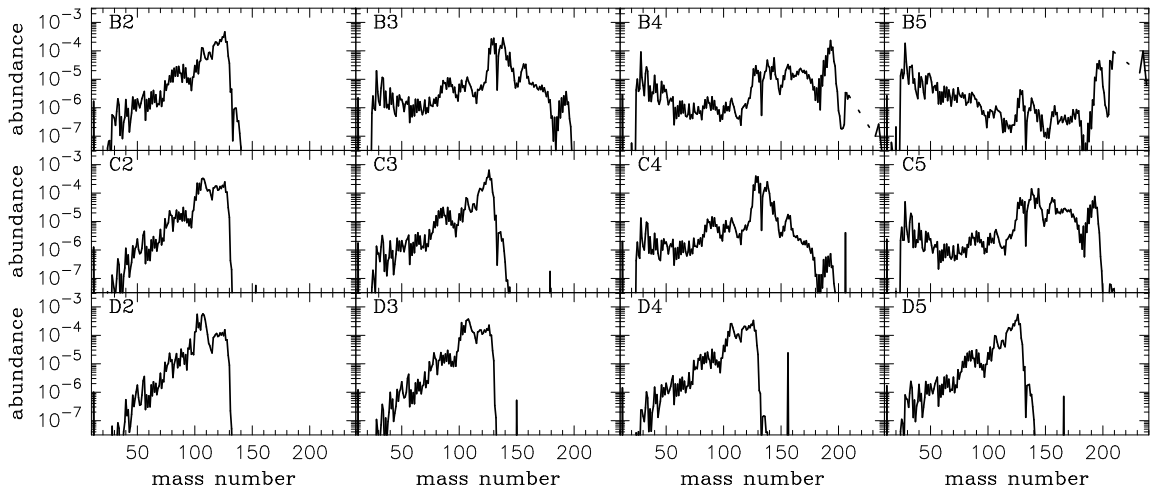}
\caption{Final abundances obtained by the nucleosynthesis calculations
for wind models listed in Table~1 (except for A1-A5) as a function of
atomic mass number.}
\end{figure}

\end{document}